\documentstyle[preprint,aps]{revtex}
\newcommand{\bff}[1]{{\mbox{\boldmath $#1$}}}
\begin{document}
\draft
\renewcommand\baselinestretch{1.32}

\title{Moments of Inertia of Nuclei in the 
Rare Earth Region: A Relativistic versus Non-Relativistic 
Investigation}
\author{A.\ V.\ Afanasjev\footnote{on leave of absence from 
the Laboratory of Radiation Physics,
Institute of Solid State Physics,
University of Latvia,
LV 2169 Salaspils, Miera srt. 31,
Latvia}, J.\ K\"onig, P.\ Ring}
\address{Physik-Department der Technischen Universit{\"a}t
M{\"u}nchen, D-85747 Garching, Germany}
\author{L.\ M.\ Robledo, J.\ L.\ Egido}
\address{Departamento de F\'{\i}sica Te\'orica C-XI,
Universidad Aut\'onoma de Madrid, E-28049, Spain}
\date{\today}
\maketitle
\begin{abstract}
A parameter free investigation of the moments of inertia 
of ground state rotational bands in well deformed rare-earth nuclei is
carried out using Cranked Relativistic Hartree-Bogoliubov 
(CRHB) and non-relativistic Cranked Hartree-Fock-Bogoliubov 
(CHFB) theories. In CRHB theory, the relativistic 
fields are determined by the non-linear Lagrangian with 
the NL1 force and the pairing interaction by the central
part of finite range Gogny D1S force. In CHFB theory, the 
properties in particle-hole and particle-particle
channels are defined solely by Gogny D1S forces. Using an 
approximate particle number projection before variation 
by means of the Lipkin Nogami method improves the agreement 
with the experimental data, especially in CRHB theory.
The effect of the particle number projection on the moments 
of inertia and pairing energies is larger in relativistic 
than in non-relativistic theory. 
\end{abstract}
\vspace{0.5cm}

\narrowtext
\section{Introduction}  

One of the oldest problems in our understanding of
collective motion of nuclei is the moments of inertia of
ground state rotational bands in well deformed nuclei. They 
depend in a very sensitive way on the collective properties 
such as deformations and on pairing correlations of these 
many-body systems. Since rotational bands have been detected 
in nuclei nearly fifty years ago and since the first
microscopic calculations of the moments of inertia by Inglis
\cite{In.54}, these quantities have been used as a testing
ground for nearly all microscopic theories of collective
motion. They describe the response of the strongly
interacting nuclear many-body system to an external
Coriolis field breaking time reversal symmetry.  They are,
therefore, in some sense comparable to the static magnetic
susceptibility in condensed matter physics. 

The earliest microscopic calculations were based on a mean 
field of a deformed harmonic oscillator \cite{In.54,BM.55,In.56}. 
In these calculations, residual interactions were
neglected. In this way one founds the values of the moments 
of inertia identical to those of a rigid body with the same
shape, in strong disagreement with the experimentally
observed values, which where considerably smaller. It has
been pointed out already very early \cite{BM.55,Mos.56}
that residual two-body interactions would lower the values
of the moment of inertia obtained in the Inglis model. The
most important correlations causing such a reduction are
pairing correlations \cite{BMP.58}.  In fact, Belyaev
\cite{Bel.59,Bel.61} showed that a simple extension of
the Inglis formula in the framework of the BCS theory is able
to reduce the theoretical moments of inertia dramatically
because of the large energy gap in the spectrum of
quasiparticle excitations occurring in the denominator of
the Belyaev formula. Therefore, the small moments of
inertia of the rotational bands provided one of the most
important experimental hints for a superfluid behaviour of
these heavy open shell nuclei.  Extended calculations 
using the theory of Belyaev have been carried out by Nilsson 
and Prior \cite{NP.61} using the BCS model based on the 
single particle spectrum of the Nilsson potential.

Apart from the fact that the results of these calculations
were relatively successful, there are, as we know today, 
a number of open problems, namely:

i) Belyaev's formula is based on generalized mean field
theory violating essential symmetries. It was pointed out
already by  Migdal \cite{Mi.59,Mi.60}, that Galileian
invariance is broken. He therefore modified the Belyaev
formula by taking into account more complicated 
correlations to correct the violation of this 
symmetry. The question of the restoration
of the broken Galileian invariance in the 
particle-hole and particle-particle channels
has been later discussed in a number of articles, see 
for example Refs.\ \cite{Bel.69,SK.90,KSKK.96} and
references therein.

ii) Since Belyaev's formula describes only quasiparticles
moving independently, higher order correlations have to be
taken into account. This has been done by Thouless and
Valatin \cite{TV.62} who considered all orders of the
interaction in a theory describing the linear response of
the system to the external Coriolis field.  Marshalek and
Weneser \cite{MW.70} showed that the method of Thouless and
Valatin preserves all the symmetries violated in the mean
field approximation in linear order. In that sense Migdal's
formula was just a special case to deal with Galilean
invariance. Marshalek showed in a series of papers (see for
example Ref.\ \cite{Mar.87}) that this is just the linear
approximation of a more general theory based on Boson
expansion techniques treating the symmetries appropriately
in all orders \cite{Mar.87}.

iii) Much more elaborated versions of the cranked Nilsson
model \cite{ALL.76,NPF.76} showed that the $\bff l^2$-term 
in this model, which corrects in an elegant way the fact that
realistic potentials for heavy nuclei are much flatter than
an oscillator in the nuclear interior, introduces a strong
spurious momentum dependence. This leads to the values for the
moments of inertia deviating considerably from the experimental 
values. However, this problem is to large extent cured 
either by Strutinsky renormalization of the moments 
of inertia \cite{ALL.76} or by an additional term to the 
cranked Nilsson potential that restores the local Galilean 
invariance \cite{BM.75,NMMS.96}.
 
 Realistic applications of the Thouless-Valatin theory are
by no means trivial. They should be based on self-consistent
solutions of the mean field equations, because only for
those solutions the RPA theory preserves the symmetries
\cite{RS.80}. In addition, they require the inversion of 
the RPA-matrix. Meyer-ter-Vehn et al.\ \cite{MSV.72} have
carried out such calculations in a restricted
configuration space replacing the self-consistent mean field
in an approximate way by the Woods-Saxon potential. As
residual interaction they used density dependent
Migdal forces $F^\omega$ in the $ph$-channel and $F^\xi$
in the $pp$-channel. These interactions have been carefully
adjusted to experimental data for the underlying
configuration space. The results of these calculations
showed, that there are indeed effects originating from both
channels, each of them modifying the Belyaev values, but
cancelling themselves to a large extent. Therefore one
could understand why older calculations \cite{NP.61}
based on the generalized mean field model gave reasonable
results as compared to the experiment.

 Nowadays there are theories available where the
Hartree-(Fock)-Bogoliubov equations can be solved in a 
fully self-consistent way in the rotating frame for 
finite angular velocity $\sl\Omega$. Using the 
resulting wavefunctions $|{\sl\Phi_\Omega}\rangle$ 
the Thouless-Valatin moment of inertia can be found 
as
\begin{eqnarray}
J~=~\left.\frac{d}{d{\sl\Omega}}
\langle{\sl\Phi_\Omega}|\hat J_x|{\sl\Phi_\Omega}\rangle
\right|_{{\sl\Omega}=0}.
\label{TW}
\end{eqnarray}
In this way one avoids the inversion of the full RPA-matrix, 
as task which is so far technically impossible for realistic 
forces in a full configuration space. 
Among these theories the properties of rotating nuclei are 
described in a way free from adjustable parameters only in 
the Cranked Relativistic Hartree-Bogoliubov (CRHB) theory 
\cite{A190,CRHB} and non-relativistic density-dependent Cranked 
Hartree-Fock-Bogoliubov (CHFB) theory with finite range Gogny 
forces \cite{ERo.93,GDBL.94}. Several realistic investigations 
of the moments of inertia in normal- and in super-deformed bands 
have been carried out in the literature in the framework of 
non-relativistic CHFB 
theory with Gogny forces \cite{ERo.93,ER.94,VER.97,VE.97,VER.99}. 
Similar investigations in the relativistic framework have 
been performed without pairing in the $A\sim 60$ \cite{A60},
80 \cite{Sr83} and $140-150$ \cite{KR.93,AKR.96,ALR.98} 
regions of superdeformation where the pairing correlations are 
expected to be considerably quenched at high spin. The recently
developed formalism of the CRHB theory has been applied so far only 
for the description of the moments of inertia in the $A\sim 190$ mass 
region of superdeformation \cite{A190,CRHB}. A very successful 
description of the moments of inertia has been obtained in the 
framework of these two theories. The aim of the present investigation 
is to find the similarities and differences between these two theories 
using in a systematic way the moments of inertia of rare-earth nuclei 
as a testing ground.

\section{Theoretical tools}

 The CRHB theory \cite{A190,CRHB} is an extension of cranked relativistic 
mean field theory \cite{KR.89,KR.93,AKR.96} to the description of 
pairing correlations in rotating nuclei. It describes the nucleus 
as a system of Dirac nucleons which interact in a relativistic
covariant manner through the exchange of virtual mesons \cite{SW.86}: 
the isoscalar scalar $\sigma$ meson, the isoscalar vector $\omega$ meson,
and the isovector vector $\rho$ meson. The phonon field $(A)$ accounts
for the electromagnetic interaction. The CRHB equations for the
fermions in the rotating frame are given in one-dimensional cranking 
approximation by 
\begin{eqnarray}
\pmatrix{ h - \Omega_x \hat{J}_x    & \hat{\Delta}  \cr
-\!\hat{\Delta}^* &     -h^* + \Omega_x \hat{J}^*_x \cr}
\pmatrix{ U_k \cr V_k } =
E_k \pmatrix{ U_k \cr V_k }
\label{CRHB}
\end{eqnarray}
where $h=h_D-\lambda$ is the Dirac Hamiltonian $h_D$ for the
nucleon with mass $m$
\begin{eqnarray}
h_D= {\bff \alpha}(-i\bff\nabla-\bff V(\bff r))~+~ V_0(\bff r)~+~\beta(m+S(\bff r))
\label{Dirac}
\end{eqnarray}
minus the chemical potential $\lambda$ defined from the average
particle number constraint
\begin{eqnarray}
\langle{\sl\Phi_\Omega}|\hat N|{\sl\Phi_\Omega}\rangle~=~N.
\end{eqnarray}
The Dirac Hamiltonian contains a repulsive vector potential
$V_0(\bff r)$, an attractive scalar potential $S(\bff r)$ and 
the magnetic potential $\bff V (\bff r)$ which leads to 
non-vanishing currents in the systems with broken 
time-reversal symmetries \cite{KR.93,AKR.96}. These 
currents play an extremely important role in the description 
of the moments of inertia \cite{KR.93,AKR.96} and thus they 
are taken into account fully self-consistently in the 
calculations. In Eq.\ 
(\ref{CRHB}), $U_k$ and $V_k$ are quasiparticle Dirac 
spinors and $E_k$ denotes the quasiparticle energies. 
Furthermore, $\hat{J}_x$ and $\Omega_x$ are the projection 
of total angular momentum on the rotation axis and the 
rotational frequency.

The time-independent inhomogeneous Klein-Gordon equations 
for the mesonic fields are given by
\begin{eqnarray}
\left\{-\Delta-({\sl\Omega}_x\hat{L}_x)^2+m_\sigma^2\right\}~
\sigma(\bff r)&=&
-g_\sigma\left[\rho_s^p(\bff r)+\rho_s^n(\bff r)\right]
-g_2\sigma^2(\bff r)-g_3\sigma^3(\bff r),
\nonumber \\
\left\{-\Delta-({\sl\Omega}_x\hat{L}_x)^2+m_\omega^2\right\}
\omega_0(\bff r)&=&
g_\omega\left[\rho_v^p(\bff r)+\rho_v^n(\bff r)\right],
\nonumber \\
\left\{-\Delta-({\sl\Omega}_x(\hat{L}_x+\hat{S}_x))^2+
m_\omega^2\right\}~
\bff\omega(\bff r)&=&
g_\omega\left[\bff j^p(\bff r)+\bff j^n(\bff r)\right],
\label{KGCRMF}
\end{eqnarray}
with source terms involving the various nucleonic densities
and currents
\begin{eqnarray}
\rho_s^i(\bff r) = \sum_{k>0} 
 [V_k^i(\bff r)]^{\dagger} \hat{\beta} V_k^i (\bff r), \qquad & &
\rho_v^i(\bff r) = \sum_{k>0} 
 [V_k^i(\bff r)]^{\dagger} V_k^i (\bff r) 
\nonumber \\
\bff j^i(\bff r) &=&  \sum_{k>0} 
[V_k^i(\bff r)]^{\dagger} \hat{\bff\alpha} V_k^i (\bff r).
\label{sourceHB}
\end{eqnarray}
  The sums over $k>0$ run over all quasiparticle states corresponding 
to positive energy single-particle states ({\it no-sea approximation})
and the indexes $i$ could be either $n$ (neutrons) or $p$ (protons). 
For simplicity, the equations for the $\rho$ meson and the Coulomb 
fields are omitted in Eqs.\ (\ref{KGCRMF}) since they have the
structure similar to the equations for the $\omega$ meson, see 
Refs.\ \cite{A190,CRHB} for details. Since the 
coupling constant of the electromagnetic interaction is small compared
with the coupling constants of the meson fields, the Coriolis term for
the Coulomb potential $A_0(\bff r)$ and the spatial components of the 
vector potential $\bff A(\bff r)$ are neglected in the calculations. 

The pairing potential $\Delta$ in Eq.\ (\ref{CRHB}) is given by
\begin{eqnarray}
\Delta \equiv \Delta_{ab}~=~\frac{1}{2}\sum_{cd} V^{pp}_{abcd} \kappa_{cd}
\label{gap}
\end{eqnarray}
where the indices $a,b,\dots$ contain the space coordinates
$\bff r$ as well as the Dirac and isospin indices $s$ and 
$t$. It contains the pairing tensor $\kappa$
\begin{eqnarray}
\kappa \equiv \kappa(\bff r,s,t,\bff r',s',t)~=~\sum_{E_k>0} 
V^*_{k}(\bff r,s,t) U_{k}(\bff r',s',t)
\label{kappa}
\end{eqnarray}
and the matrix elements $V^{pp}_{abcd}$ of the effective
interaction in the $pp$-channel. In the present version 
of CRHB theory, pairing correlations are only considered 
between the baryons, because pairing is a genuine non-relativistic 
effect, which plays a role only in the vicinity of the Fermi 
surface. The central part of the Gogny interaction containing
two different finite range terms (see Eq.\ (\ref{Gogny})) is
employed in the $pp$ (pairing) channel.

   The CRHB calculations have been performed with the NL1
parametrization \cite{NL1} of the RMF Lagrangian. The D1S
set of parameters \cite{BGG.84} is used for the Gogny force in 
the pairing channel. The CRHB-equations are solved in the basis 
of an anisotropic three-dimensional harmonic oscillator in 
Cartesian coordinates. A basis deformation of $\beta_0=0.3$ has 
been used. All fermionic and bosonic states belonging to the 
shells up to $N_F=13$ and $N_B=16$ are taken into account in 
the diagonalisation and the matrix inversion, respectively. 
This truncation scheme provides reasonable numerical accuracy 
for the physical observables which as estimated in the 
calculations with larger fermionic basis is on the level of 
$\sim 1.5\%$ or better for kinematic moment of inertia $J^{(1)}$ 
and charge quadrupole moments $Q_0$. In order to calculate the
derivative with respect to $\Omega$ in Eq.\ (\ref{TW}), all CRHB 
calculations have been performed at rotational frequency 
$\Omega_x=0.05$ MeV. 

 The starting point of the non-relativistic CHFB theory based 
on the Gogny force is the phenomenological finite range 
two-body interaction of the form \cite{VER.99}
\begin{eqnarray}
V^{pp}(1,2) & = & \sum_{i=1,2} e^{-[({\bff r}_1-{\bff r} _2)/\mu_i]^2} 
(W_i+B_i P^{\sigma}- H_i P^{\tau} - M_i P^{\sigma}
P^{\tau}) + \nonumber \\
&+& i W_{LS} (\bff\nabla_{12} \wedge \delta(\bff r_1 -\bff r_2)  
\bff\nabla_{12}) (\bff\sigma_1+\bff\sigma_2) \nonumber \\
&+& t_3 (1+P^{\sigma} x_0) \delta(\bff r_1-\bff r_2)
[\rho(\bff R)]^{1/3}
\label{Gogny}
\end{eqnarray}
which is used simultaneously both in $pp$ and $ph$ channels.
In Eq.\ (\ref{Gogny}), $\bff R=(\bff r_1+\bff r_2)/2$. The transformation 
to the rotating frame \cite{VER.99} leads 
to equations similar to (\ref{CRHB},\ref{gap},\ref{kappa}).
The only difference is that the Dirac Hamiltonian of Eq.\ 
(\ref{Dirac}) is replaced by the non-relativistic Hartree-Fock
Hamiltonian $h_{ij}$ containing the density-dependent Gogny 
force and the rearrangement term $\partial \Gamma_{ij}$, 
stemming from the density dependence of the force
\begin{eqnarray}
h_{ij}  \rightarrow  h_{ij}+\partial \Gamma_{ij}
=t_{ij}+\sum_{qq^{\prime }}\widetilde{\upsilon }_{iqjq^{\prime }}
\rho _{q^{\prime }q} 
 +  \left\langle \frac{\delta H^{\prime }}{\delta
\rho } f_{ij}({\bff R})\right\rangle
\end{eqnarray}
In the above expression $f_{ij}({\bff r})$ is the quantity appearing 
in the second quantization form of the density operator 
$\rho ({\bff r})=\sum_{ij} f_{ij}({\bff r}) c^+_i c_j$.
The parameter set D1S \cite{BGG.84} has been used in the 
present calculations. The CHFB-equations are again solved in 
the basis of an anisotropic three-dimensional harmonic 
oscillator in Cartesian coordinates with the oscillator 
length $b_0=1.98$ fm and the deformation of basis $\beta_0=0.3$. 
Only single-particle states satisfying the condition
\begin{eqnarray}
\hbar\omega_x n_x + \hbar\omega_y n_y +
\hbar\omega_z n_z \le N_{max}\hbar\omega_0\,\,\,\,\,\,\,\,
{\rm with}\,\,\,N_{max}=11.1
\end{eqnarray}
have been included in the basis. The HFB equation has been solved 
with the Conjugated Gradient Method \cite{CG.95}.

   We also consider in this investigation the fluctuations 
in the pairing field by using the technique of an approximate
particle number projection before the variation introduced 
by Lipkin and Nogami (further APNP(LN)) and discussed in 
detail in non-relativistic case in Refs.\ 
\cite{SWM.94,GBDFH.94,VER.99}. In the relativistic case, the 
same approximate particle number projection is used but only 
the $pp$-part of the interaction is taken into account for 
the Lipkin-Nogami procedure, see Ref.\ \cite{CRHB} for 
details. 

\section{Results and discussion}

In order to see the dependence of the results on proton and 
neutron numbers several nuclei in the Gd, Dy, Er and Yb 
isotope chains, the ground state rotational bands of which 
are close to rotational limit ($E(4^+)/E(2^+)\approx 3.3$), 
have been selected for the present study. The results of 
relativistic and non-relativistic calculations with and without 
APNP(LN) are presented in Tables 1-3 and Figs. 1-2 and compared 
with the experiment. Such quantities as charge
quadrupole moments [deformations], pairing 
energies and moments of inertia are discussed below. 

 The calculated and experimental charge
quadrupole moments $Q$ and quadrupole deformation 
parameters $\beta$ derived from $Q$ by
\begin{eqnarray}
Q~=~\sqrt{\frac{16\pi}{5}} \frac{3}{4\pi} Z R_0^2\beta 
\,\,\,\,\,\,{\rm where}\,\,\, R_0=1.2 A^{1/3}
\end{eqnarray}
are shown in Table 1 and Fig. 1. The general feature is that
the charge quadrupole moments $Q$
obtained in relativistic calculations are
larger than the ones of the non-relativistic calculations.
In the non-relativistic case, the $Q$ values
calculated
with APNP(LN) are slightly smaller than the ones obtained 
without APNP(LN) because of the larger pairing correlations 
(see Table 2), which in general favours more spherical 
configurations. This trend also persists in the relativistic 
case, but there are the cases ($^{164,166}$Er, $^{164,166}$Yb)
in which APNP(LN) leads to larger charge quadrupole moments 
as compared with 
unprojected calculations. As shown in Fig. 1, the non-relativistic 
results are somewhat closer to the experiment  than the relativistic 
ones. However, considering that the experimental values of $Q$
are subject of considerable experimental errors \cite{Ram.87}, 
one can 
conclude that both theories  describe experimental charge
quadrupole moments
reasonable well which allows us to proceed further with the 
study of more sensitive quantities such as moments of inertia.
 
 In Hartree-(Fock)-Bogoliubov calculations the size of the 
pairing correlations is usually measured in terms of the 
pairing energy defined as
\begin{eqnarray}
E_{pair}~=~-\frac{1}{2}\mbox{Tr} (\Delta\kappa).
\end{eqnarray}
This is not an experimentally accessible quantity, but it is a 
measure for the size of the pairing correlations in the theoretical 
calculations. These quantities are shown in Table 2 and Fig.\ 2 for 
protons and neutrons separately. Both in relativistic and non-relativistic 
calculations, we observe that APNP(LN) leads to an increase of 
the pairing energies. This increase shows large variations as a 
function of the proton and neutron numbers. In general, this 
increase is larger in relativistic calculations. For example, 
proton pairing energies increase on average by factor 2.16 
(with minimal and maximal increase being equal to 1.53 
($^{156}$Dy) and 3.36 ($^{160}$Gd)). In non-relativistic 
calculations, the average increase of proton pairing energies 
is only 1.55 (with minimal and maximal increase being equal 
to 1.11 ($^{156}$Dy) and 1.97 ($^{170}$Yb)). Neutron pairing 
energies behave in a similar way but there the difference between 
relativistic and non-relativistic calculations is smaller: the 
average increase of neutron pairing energies due to APNP(LN) is 
1.81 in the relativistic 
and 1.73 in the non-relativistic calculations. In some cases, 
as for example in the Gd isotopes, the increase of neutron 
pairing energies due to APNP(LN) is larger in non-relativistic 
calculations. This increase of pairing energies due to APNP(LN) 
will lead to an increase of the pairing gaps, as it is well 
known from many phenomenological calculations using the 
monopole pairing force \cite{ER.82a,ER.82b}. We also see 
that with few exceptions the pairing energies are smaller 
in relativistic calculations. An additional effect of APNP(LN) 
is the increase of absolute values of binding energies. In 
relativistic calculations, APNP(LN) provides an additional 
binding by $\approx -2.5$ MeV.  

 Calculated moments of inertia are given in Table 3 and Fig. 1. 
Comparing the results of calculations without APNP(LN), it is 
clear that the moments of inertia are systematically larger in the 
relativistic case. Although one cannot completely exclude that 
this feature is to some extent connected with a different angular 
momentum content of single-particle orbitals in relativistic 
and non-relativistic calculations, a detailed analysis of pairing
energies and moments of inertia suggests that this fact can be
explained in a more realistic way by the different effective masses 
of the two theories: $m^*/m\sim 0.6$ in RMF theory and $\sim 0.7$ 
in the non-relativistic theory. Thus the corresponding level density 
in the vicinity of the Fermi level is smaller in the 
relativistic theory which in general leads to weaker pairing 
correlations (see Table 2 and discussion above) as compared 
with non-relativistic calculations and as a result to larger 
moments  of inertia. APNP(LN) restores to large extent the 
correct size of pairing correlations and thus its effect 
is larger in relativistic calculations. The average decrease 
of the moments of inertia due to APNP(LN) over the considered 
set of nuclei is 1.35 and 1.15 in relativistic and non-relativistic 
calculations, respectively. It is also clearly seen that APNP(LN) 
improves in average and especially in relativistic case the 
agreement between experimental and calculated moments of 
inertia. The level of agreement between calculations with
APNP(LN) and experiment is similar in both theories, however, 
some discrepancies still remain.

\section{Conclusions}

  In conclusion, the moments of inertia, charge quadrupole
moments and pairing energies of well-deformed nuclei in the 
rare-earth region have been investigated within relativistic and 
non-relativistic mean field theories with and without 
approximate particle number projection by means of the 
Lipkin-Nogami method. With no adjustable parameters it
was possible to obtain good description of experimental 
data. It was found that the particle number projection 
plays a more important role in the relativistic calculations 
most likely reflecting the lower effective mass. In addition,
it has larger impact on the moments of inertia and the pairing
energies as compared with the charge quadrupole moments.
Remaining 
deviations from experimental data could be related either 
to the parametrization of the mean field or to the interaction 
in the pairing channel or to the approximate character of 
the particle number projection. Further and more systematic 
investigations are needed for clarification of the main 
source of discrepancies between theory and experiment.

\vspace{0.5cm}

A.V.A. acknowledges support from the Alexander von Humboldt
Foundation. This work has been supported in part by DGICyT, 
Spain under project PB97-0023 and the Bundesministerium f\"ur
Bildung und Forschung under the project 06 TM 875. P.R. wishes 
to express his gratitude to the Spanish Ministry for Education 
for support of his work at the Universidad Aut\'onoma de Madrid in the 
framework of the A.  von Humboldt - J.C.  Mutis program for 
the Scientific Cooperation between Spain and Germany.

\section{Figure caption}

Fig. 1. Experimental and calculated charge quadrupole 
moments $Q$  
(top panels) and moments of inertia $2 J^{(1)}$ 
(bottom panels) of well deformed rare-earth nuclei. The
experimental data are shown by solid unlinked circles. The 
results of calculations with APNP(LN) are shown by the lines 
without symbols. The lines with open symbols are used 
to indicate the results of calculations without APNP(LN).
Solid and dashed lines are used for relativistic and 
non-relativistic results, respectively. 
\\
\\
Fig. 2. Calculated neutron (top panels) and proton (bottom
panels) pairing energies. The results of calculations with 
(without) APNP(LN) are shown by the lines without (with) 
symbols. Solid and dashed lines are used for relativistic 
and non-relativistic results, respectively.

\renewcommand{\arraystretch}{1.0}
\begin{table}[t]
\caption{\protect\small\label{T1}
The calculated and experimental charge quadrupole 
moments $Q$ and quadrupole deformation parameters 
$\beta$ [shown in squared brackets] for typical 
well deformed nuclei in the rare earth region.
The results of relativistic calculations are indicated by
'CRHB', while the results of non-relativistic calculations
by 'Gogny'. The calculations without and with approximate
particle number projection by means of the Lipkin-Nogami
method are shown in columns marked by 'without projection' 
and 'with projection', respectively. The experimental 
data are taken from Ref.\ \protect\cite{Ram.87}.}
\vspace{0.5cm}
 \centering
\begin{tabular}{cccccccc}
   &     &\multicolumn{2}{c}{without projection} & 
          \multicolumn{2}{c}{with projection}    &  Expt. \\
\hline
   & A   & CRHB         & Gogny &   CRHB       & Gogny &        \\
\hline
Gd & 154 & 6.715~[0.335]  & 6.074~[0.303]  & 5.907~[0.295]  & 5.606~[0.280] & 6.221~[0.310]  \\
   & 156 & 7.199~[0.356]  & 6.792~[0.336]  & 6.886~[0.341]  & 6.601~[0.327] & 6.830~[0.338]  \\
   & 158 & 7.383~[0.362]  & 7.077~[0.347]  & 7.262~[0.356]  & 6.961~[0.341] & 7.104~[0.348]  \\
   & 160 & 7.577~[0.369]  & 7.286~[0.354]  & 7.490~[0.364]  & 7.200~[0.350] & 7.265~[0.353]  \\
   &     &                &                &                &               &        \\
Dy & 156 & 5.860~[0.281]  & 5.994~[0.287]  & 5.610~[0.269]  & 5.438~[0.261] & 6.107~[0.293] \\
   & 158 & 7.032~[0.334]  & 6.990~[0.333]  & 6.711~[0.319]  & 6.538~[0.311] & 6.844~[0.326] \\
   & 160 & 7.496~[0.354]  & 7.297~[0.344]  & 7.373~[0.348]  & 7.102~[0.335] & 7.13~[0.337] \\
   & 162 & 7.711~[0.361]  & 7.492~[0.350]  & 7.697~[0.360]  & 7.382~[0.345] & 7.28~[0.341] \\
   & 164 & 7.928~[0.368]  & 7.626~[0.354]  & 7.883~[0.366]  & 7.543~[0.350] & 7.503~[0.348] \\
   &     &                &                &                &               &       \\
Er & 164 & 7.671~[0.345]  & 7.585~[0.341]  & 7.791~[0.351]  & 7.522~[0.339] & 7.402~[0.333] \\
   & 166 & 8.047~[0.359]  & 7.781~[0.347]  & 8.075~[0.361]  & 7.728~[0.345] & 7.656~[0.342] \\
   & 168 & 8.213~[0.364]  & 7.838~[0.347]  & 8.151~[0.361]  & 7.831~[0.347] & 7.63~[0.338] \\
   & 170 & 8.137~[0.358]  & 7.899~[0.347]  & 8.075~[0.355]  & 7.782~[0.342] & 7.65~[0.336] \\
   &     &                &                &                &               &              \\
Yb & 164 & 6.552~[0.287]  & 6.900~[0.302]  & 6.602~[0.289]  & 6.828~[0.299] & 6.60~[0.289] \\
   & 166 & 7.339~[0.318]  & 7.594~[0.330]  & 7.653~[0.332]  & 7.508~[0.326] & 7.19~[0.312] \\
   & 168 & 8.362~[0.360]  & 7.997~[0.344]  & 8.222~[0.354]  & 7.864~[0.339] & 7.59~[0.327] \\
   & 170 & 8.546~[0.365]  & 8.147~[0.348]  & 8.354~[0.357]  & 8.013~[0.342] & 7.57~[0.324] \\
\end{tabular}
\end{table}  
\renewcommand{\arraystretch}{1.0}

\newpage

\renewcommand{\arraystretch}{1.0}
\begin{table}[t] 
\caption{\protect\small\label{T2}
Pairing energies for typical well deformed nuclei in 
the rare earth region. For details of this table see 
caption of Table 1.}
\vspace{0.5cm}
\centering
\begin{tabular}{crrrrrrrrr}
   & A   &\multicolumn{4}{c}{$E^n_{pair}$}  & 
          \multicolumn{4}{c}{$E^p_{pair}$} \\ 
\hline
   &     &\multicolumn{2}{c}{without projection}   & 
          \multicolumn{2}{c}{with projection}      &
          \multicolumn{2}{c}{without projection}   & 
          \multicolumn{2}{c}{with projection}      \\
\hline
   & A   & CRHB      & Gogny  & CRHB  & Gogny &                   CRHB      & Gogny  & CRHB  & Gogny \\
\hline
Gd & 154 & ~~-6.790 & -7.413 & -11.264 &   -14.566 &            ~~-4.556 & -8.857 & -10.176 & -10.828  \\
   & 156 & ~~-6.039 & -6.236 & -11.552 &   -12.871 &            ~~-3.635 & -7.740 & -8.977  & -10.778  \\
   & 158 & ~~-7.071 & -6.511 & -11.495 &   -12.355 &            ~~-3.151 & -7.247 & -8.748  & -10.487  \\
   & 160 & ~~-7.174 & -6.731 & -11.385 &   -11.961 &            ~~-2.579 & -6.830 & -8.686  & -10.341  \\
   &     &          &        &         &              &&&&\\
Dy & 156 & ~~-5.965 & -8.017 & -10.904 &   -15.610  &           ~~-7.293 & -9.849 & -11.167 & -10.888 \\
   & 158 & ~~-7.007 & -8.085 & -11.742 &   -13.870  &           ~~-5.596 & -7.773 & -9.888  & -11.234 \\
   & 160 & ~~-7.706 & -8.067 & -11.712 &   -13.101  &           ~~-3.077 & -6.980 & -9.470  & -11.090 \\
   & 162 & ~~-6.331 & -7.887 & -11.440 &   -12.581  &           ~~-2.859 & -6.323 & -9.289  & -10.837 \\
   & 164 & ~~-5.237 & -6.755 & -11.046 &   -12.169  &           ~~-3.497 & -5.952 & -9.187  & -10.332 \\
   &     &          &        &         &     &&&&\\
Er & 164 & ~~-6.154 & -8.686 & -11.562 & -13.157  &           ~~-5.924 & -7.176 & -9.938  &  -11.256  \\
   & 166 & ~~-6.549 & -7.520 & -11.052 & -12.621  &           ~~-6.019 & -6.213 & -9.701  &  -10.603  \\
   & 168 & ~~-4.036 & -5.569 & -10.784 & -12.043  &           ~~-5.695 & -5.555 & -9.492  &  -9.904   \\
   & 170 & ~~-5.741 & -4.881 & -10.998 & -11.58   &           ~~-4.968 & -5.157 & -9.289  &  -9.464   \\
   &     &          &        &         &     &&&&\\
Yb & 164 & ~~-8.651 & -9.462 & -12.069 &  -11.256  &            ~~-6.388 & -9.336 & -10.435 & -11.708 \\
   & 166 & ~~-7.919 & -9.114 & -11.527 &  -10.603  &           ~~-6.459 & -8.223 & -10.427  & -11.350 \\
   & 168 & ~~-6.702 & -7.947 & -10.946 &  -9.904   &             ~~-6.024 & -6.715 & -10.12 & -10.771 \\
   & 170 & ~~-4.353 & -6.061 & -10.603 &  -9.464   &           ~~-5.274 & -5.105 &  -9.795  & -10.051 \\
\end{tabular}
\end{table}  
\renewcommand{\arraystretch}{1.0}

\newpage
\renewcommand{\arraystretch}{1.0}
\begin{table}[t]
\caption{\protect\small\label{T3}
Moments of inertia 2$J^{(1)}$ in units of MeV$^{-1}$
for typical well deformed nuclei in the rare earth region.
The experimental values are extracted from the energies
of the first excited $2^+$ states given in Ref.\ 
\protect\cite{Ram.87}. For other details of table see 
caption of Table 1.}
\vspace{0.5cm}
\centering
\begin{tabular}{crrrrrr}
   &     &\multicolumn{2}{c}{without projection} & 
          \multicolumn{2}{c}{with projection} & expt. \\
\hline
   & A   & RHB        & Gogny & RHB       & Gogny &    \\
\hline
Gd & 154 & ~~~~~78.19 & 64.79 &~~~~~49.26 & 48.56 &~~~~~48.75 \\
   & 156 & ~~~~~88.79 & 78.74 &~~~~~62.15 & 67.96 &~~~~~67.44 \\
   & 158 & ~~~~~86.38 & 79.05 &~~~~~66.42 & 70.90 &~~~~~75.46 \\
   & 160 & ~~~~~87.30 & 81.56 &~~~~~68.80 & 66.50 &~~~~~79.72 \\
   &     &            &       &           &       &           \\
Dy & 156 & ~~~~~63.22 & 57.52 &~~~~~43.32 & 47.79 &~~~~~43.52 \\
   & 158 & ~~~~~80.45 & 72.81 &~~~~~56.17 & 65.66 &~~~~~60.64 \\
   & 160 & ~~~~~93.76 & 76.14 &~~~~~63.71 & 67.38 &~~~~~69.13 \\
   & 162 & ~~~~~98.22 & 77.56 &~~~~~67.24 & 68.00 &~~~~~74.38 \\
   & 164 & ~~~~~99.92 & 82.41 &~~~~~69.80 & 70.69 &~~~~~81.75 \\
   &     &            &       &           &  &           \\
Er & 164 & ~~~~~88.49 & 72.50 &~~~~~64.73 & 64.62 &~~~~~65.65 \\
   & 166 & ~~~~~83.34 & 76.40 &~~~~~68.20 & 69.40 &~~~~~74.46 \\
   & 168 & ~~~~~93.74 & 82.19 &~~~~~69.04 & 69.23 &~~~~~75.19 \\
   & 170 & ~~~~~82.30 & 81.08 &~~~~~67.39 & 71.55 &~~~~~76.25 \\
   &     &            &       &           &  &           \\
Yb & 164 & ~~~~~61.66 & 60.72 &~~~~~49.88 & 57.10 &~~~~~48.66 \\
   & 166 & ~~~~~70.54 & 67.72 &~~~~~61.24 & 62.44 &~~~~~58.61 \\
   & 168 & ~~~~~82.96 & 75.00 &~~~~~67.37 & 68.91 &~~~~~68.39 \\
   & 170 & ~~~~~93.12 & 80.53 &~~~~~68.61 & 71.69 &~~~~~71.21 \\
\end{tabular}
\end{table}  
\renewcommand{\arraystretch}{1.0}

\newpage
\begin{thebibliography}{999}

\bibitem{In.54} D.\ R.\ Inglis, Phys.\ Rev. {\bf 96}, 1059 (1954).

\bibitem{BM.55} A.\ Bohr and B.\ R.\ Mottelson, 
         Mat.\ Fys.\ Medd.\ Dan.\ Vid.\ Selsk. {\bf 30}, No.\ 1 (1955).

\bibitem{In.56} D.\ R.\ Inglis, Phys.\ Rev. {\bf 103}, 1786 (1956).

\bibitem{Mos.56} S.\ A.\ Moszkowski, Phys.\ Rev. {\bf 103}, 1328 (1956). 

\bibitem{BMP.58} A.\ Bohr, B.\ R.\ Mottelson, and D.\ Pines, 
                 Phys.\ Rev. {\bf 110}, 936 (1958).

\bibitem{Bel.59} S.\ T.\ Belyaev, Mat.\ Fys.\ Medd.\ Dan.\ Vid.\ Selsk. 
        {\bf 31}, No.\ 11 (1959).

\bibitem{Bel.61} S.\ T.\ Belyaev, Nucl.\ Phys. {\bf 24}, 322 (1961).

\bibitem{NP.61} S.\ G.\ Nilsson and O.\ Prior, 
       Mat.\ Fys.\ Medd.\ Dan.\ Vid.\ Selsk. {\bf 32}, No.\ 16 (1961).
 
\bibitem{Mi.59} A.\ B.\ Migdal, Nucl.\ Phys. {\bf 13}, 655 (1959).

\bibitem{Mi.60} A.\ B.\ Migdal, Sov.\ Phys.\ JETP {\bf 10}, 176 (1960).

\bibitem{Bel.69} 
S.\ T.\ Belyaev, Phys.\ Lett. {\bf 28B}, 365 (1969).

\bibitem{SK.90} 
H.\ Sakamoto and T.\ Kishimoto, Phys.\ Lett. {\bf B 245}, 321 (1990).

\bibitem{KSKK.96} 
T.\ Kubo, H.\ Sakamoto, T.\ Kammuri, and T.\ Kishimoto, 
Phys.\ Rev. C {\bf 54}, 2331 (1996).

\bibitem{TV.62} D.\ J.\ Thouless and J.\ G.\ Valatin, 
                Nucl.\ Phys. {\bf 31}, 211 (1962).

\bibitem{MW.70} E.\ R.\ Marshalek and J.\ Weneser, 
                Phys.\ Rev. C {\bf 2}, 1682 (1970).

\bibitem{Mar.87} 
E.\ R.\ Marshalek, Phys.\ Rev. C {\bf 35}, 1900 (1987).

\bibitem{ALL.76} G.\ Andersson, S.\ E.\ Larsson, G.\ Leander, 
                 P.\ M\"oller, S.\ G.\ Nilsson, I.\ Ragnarsson, 
                 S.\ \AA berg, R.\ Bengtsson, J.\ Dudek, 
                 B.\ Nerlo-Pomorska, K.\ Pomorski, and Z.\ Szyma{\'n}ski, 
                 Nucl.\ Phys. {\bf A268}, 205 (1976).

\bibitem{NPF.76} K.\ Neerg{\aa}rd, V.\ V.\ Pashkevich, and 
                 S.\ Frauendorf, Nucl.\ Phys. {\bf A262}, 61 (1976).

\bibitem{BM.75} A.\ Bohr and B.\ R.\ Mottelson,
{\it Nuclear Structure}, Vol.\ 2 (Benjamin, New York, 
1975)

\bibitem{NMMS.96} T.\ Nakatsukasa, K.\ Matsuyanagi, S.\ Mizutori, 
and Y.\ R.\ Shimizu, Phys.\ Rev. C {\bf 53}, 2213 (1996).

\bibitem{RS.80} P.\ Ring and P.\ Schuck, 
                {\it The Nuclear Many-body Problem} 
                (Springer Verlag, Heidelberg, 1980) 

\bibitem{MSV.72} J.\ Meyer, J.\ Speth, and J.\ H.\ Vogeler, 
                 Nucl.\ Phys. {\bf A193}, 60 (1972).

\bibitem{A190} A.\ V.\ Afanasjev, J.\ K\"onig, and P.\ Ring,
               Phys.\ Rev. C {\bf 60}, 051303 (1999). 

\bibitem{CRHB} A.\ V.\ Afanasjev, P.\ Ring, and J.\ K\"onig,
Nucl.\ Phys. A, in press (see also report nucl-th/0001054) 

\bibitem{ERo.93} J.\ L.\ Egido and L.\ Robledo, 
                  Phys.\ Rev.\ Lett. {\bf 70}, 2876 (1993).

\bibitem{GDBL.94} M.\ Girod, J.\ P.\ Delaroche, J.\ F.\ Berger
               and J.\ Libert, Phys.\ Lett. {\bf B 325}, 1 (1994).

\bibitem{ER.94} J.\ L.\ Egido and L.\ M.\ Robledo, 
                Nucl.\ Phys. {\bf A570}, 69c (1994).

\bibitem{VER.97} A.\ Valor, J.\ L.\ Egido and 
                 L.\ M.\ Robledo, Phys.\ Lett. {\bf B 392}, 249 (1997).

\bibitem{VE.97} A.\ Villafranca and J.\ L.\ Egido,
                Phys.\ Lett. {\bf B 408}, 35 (1997).

\bibitem{VER.99} A.\ Valor, J.\ L.\ Egido and 
                 L.\ M.\ Robledo, Nucl.\ Phys. {\bf A665}, 
                 46 (2000)

\bibitem{A60} A.\ V.\ Afanasjev, I.\ Ragnarsson, and 
              P.\ Ring, Phys.\ Rev. C {\bf 59}, 3166 (1999).

\bibitem{Sr83} A.\ V.\ Afanasjev, J.\ K\"onig, and P.\ Ring,
               Phys.\ Lett. {\bf B367}, 11 (1996).

\bibitem{KR.93} J.\ K{\"o}nig and P.\ Ring,
                Phys.\ Rev.\ Lett. {\bf 71}, 3079 (1993).

\bibitem{AKR.96} A.\ V.\ Afanasjev, J.\ K{\"o}nig, and 
                P.\ Ring, Nucl.\ Phys. {\bf A608}, 107 (1996).

\bibitem{ALR.98} A.\ V.\ Afanasjev, G.\ A.\ Lalazissis, 
                 and P.\ Ring, Nucl.\ Phys. {\bf A634}, 395 (1998).

\bibitem{KR.89} W.\ Koepf and P.\ Ring, 
                Nucl.\ Phys. {\bf A493}, 61 (1989).

\bibitem{SW.86} B.\ D.\ Serot and J.\ D.\ Walecka, 
                Adv.\ Nucl.\ Phys. {\bf 16}, 1 (1986).

\bibitem{NL1}  P.\ G.\ Reinhard, M.\ Rufa, J.\ Maruhn, 
               W.\ Greiner, and J.\ Friedrich,
               Z.\ Phys. {\bf A323}, 13 (1986).

\bibitem{BGG.84} J.\ F.\ Berger, M.\ Girod, and D.\ Gogny, 
                 Nucl.\ Phys. {\bf A428}, 23c (1984).

\bibitem{CG.95} J.\ L.\ Egido, J.\ Lessing, V.\ Martin, and L.\ M.\ Robledo,
                Nucl.\ Phys. {\bf A594}, 70 (1995).

\bibitem{GBDFH.94} B.\ Gall, P.\ Bonche, J.\ Dobaczewski,
H.\ Flocard, and P.-H.\ Heenen, Z.\ Phys. {\bf A 348}, 183 (1994).

\bibitem{SWM.94} W.\ Satu{\l}a, R.\ Wyss, and P.\ Magierski,
Nucl.\ Phys. {\bf A578}, 45 (1994).

\bibitem{Ram.87} S.\ Raman, C.\ H.\ Malarkey, W.\ T.\ Milner, 
                 C.\ W.\ Nestor, JR., and P.\ H.\ Stelson, 
                 At.\ Data Nucl.\ Data Tables {\bf 36}, 1 (1987).

\bibitem{ER.82a} J.\ L.\ Egido and P.\ Ring, 
                 Nucl.\ Phys. {\bf A383}, 189 (1982).

\bibitem{ER.82b} J.\ L.\ Egido and P.\ Ring, 
                 Nucl.\ Phys. {\bf A388}, 19 (1982).

\end{thebibliography}
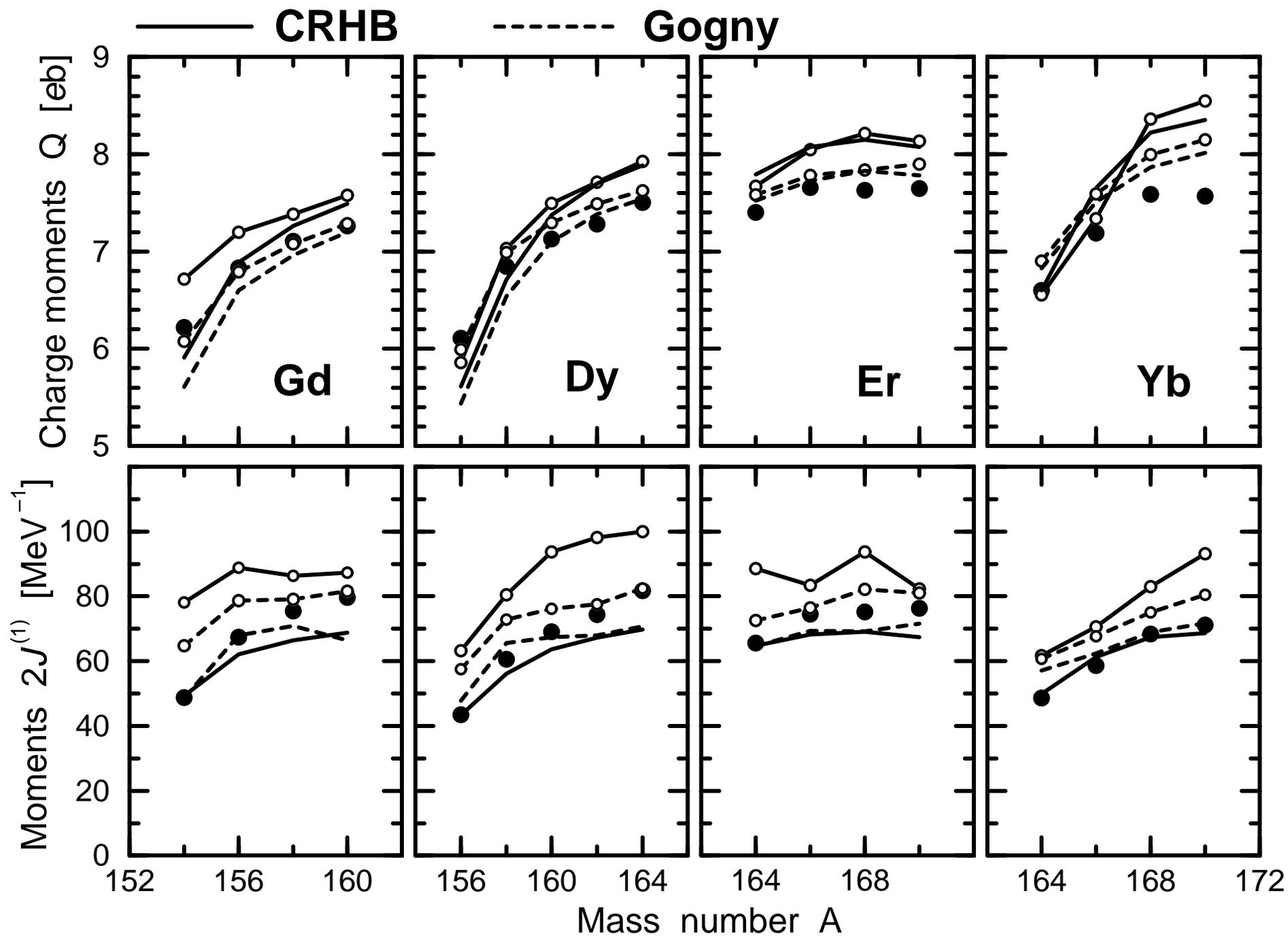
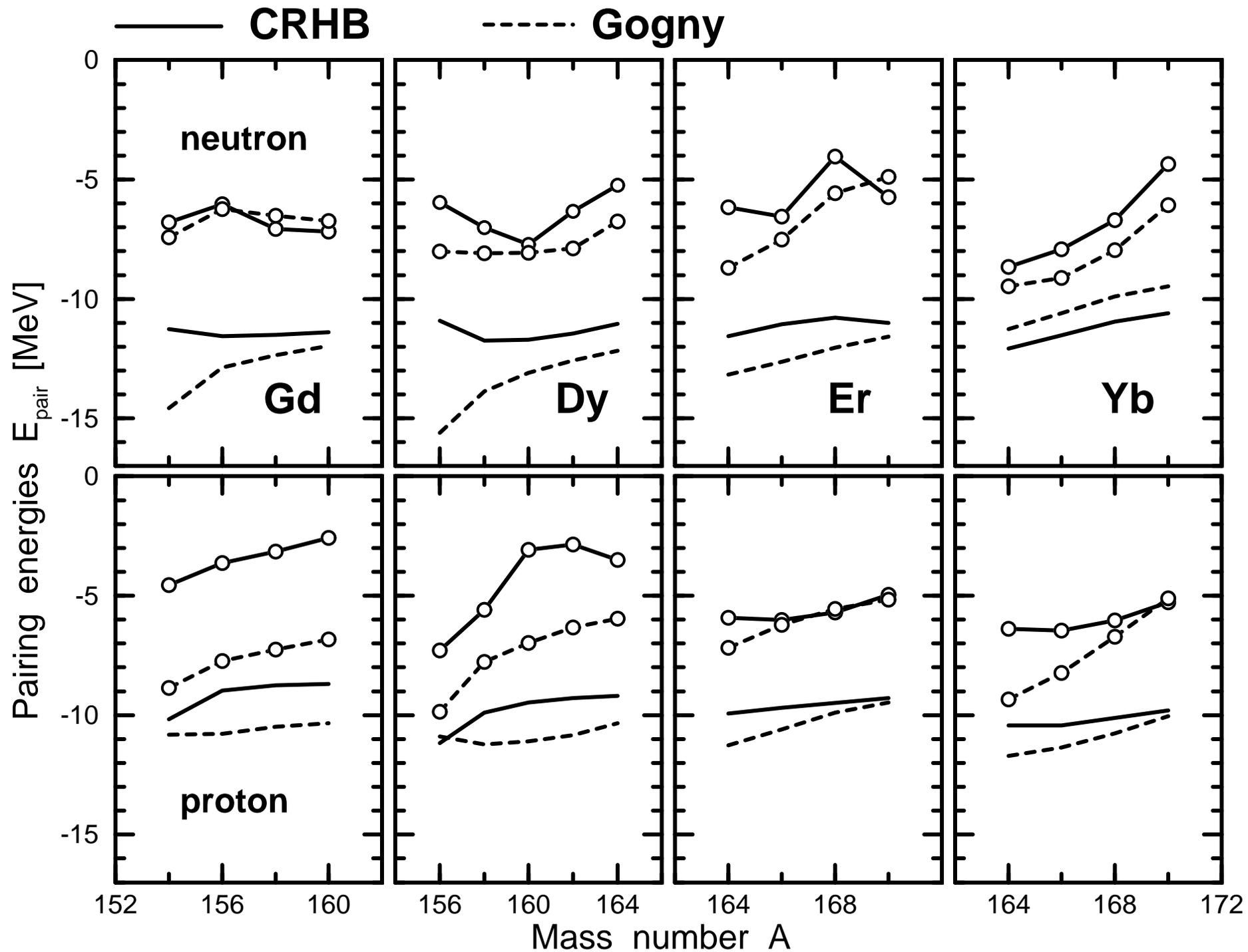
\end{document}